\documentclass{article}

\begin{document}

\title{A \bigskip NEW 2D INTEGRABLE SYSTEM WITH A QUARTIC SECOND INVARIANT}
\author{Yehia Hamad M. \\
Department of Mathematics, Faculty of Science,\\
Mansoura University, Mansoura 35516, Egypt\\
Email: hyehia@mans.edu.eg}
\maketitle

\begin{abstract}
The construction of all 2-D Lagrangian systems, which admits besides the
energy another integral of motion that is quartic in velocities was reduced
in our previous article (J. Phys. A: Math. Gen., 39, 5807--5824, 2006) to a
single nonlinear PDE.

In the present note we introduce a new solution of this equation, leading to
a new integrable system with a quartic integral, which involves 16 free
parameters. A special case of the new system admits interpretation in a
problem of rigid body dynamics. It gives a new integrable variation of the
cases due to Kovalevskaya (1889), Chaplygin (1903) Goriatchev (1916) and
Yehia (2006).
\end{abstract}

\section{Introduction}

\subsection{Historical}

The famous Kovalevskaya's integrable case of rigid body dynamics was the
first example of a mechanical system that admits an integral of motion
quartic in velocities \cite{kov}. For more than a century this case
attracted attention of many specialists, who treated explicit solution in
terms of time and gave several modifications and generalizations. Only in
the last three decades there appeared a few new integrable systems with a
quartic integral, but mainly concerning the motion of a particle in the
Euclidean plane under the action of certain potential forces. A short, but
nearly complete up to its date, list of those systems can be found in
Hietarinta's review paper \cite{Hiet}. A few cases of the same type were
obtained later in \cite{Karlovini 2002}.

In virtue of Maupertuis principle, the motion of a natural mechanical system
can be brought into equivalence (in the orbital sense) with the geodesic
flow on some Riemannian metric. Metrics on the Riemannian sphere associated
with known integrable cases of rigid body dynamics were constructed in \cite%
{Bol}. Two families of integrable systems with a quartic integral on $S^{2}$
were obtained in \cite{sel} and \cite{had-sel}. Few more works discussed
possible integrable systems with low-degree polynomials on $S^{2}$ and the
hyperbolic plane $H^{2}$ (see e.g. \cite{dul}-\cite{ran1}).

The method introduced in our work \cite{Yehia 86} and developed in several
later works, has led to construction of a large number of several-parameter
families of integrable systems with a complementary integral ranging from
second to fourth degree (see e.g. \cite{yjmp}, \cite{yjpacubic}). This
method leads in a natural way to several-parameter integrable systems that
allow as special cases, for particular values of the parameters,
interpretations as motion on various flat and curved 2-D manifolds. Although
our primary interest is in systems on Riemannian manifolds, some of the
constructed integrable systems live on pseudo-Euclidean or pseudo-Riemannian
manifolds. New examples of integrable rigid body dynamics are common
by-products of this method \cite{Yehia 86}-\cite{ym11}.

The culmination of this method was the construction of the so-called
"master" system with a quartic integral \cite{yjpamast}. It involves the
largest ever number of 21 parameters and covers almost all systems of that
type that were known earlier. Applied to rigid body dynamics, this system
gave rise to 4 new integrable cases of motion under action of potential
forces. In two cases the potential is a single valued function on the
configuration space, but multivalued in the other two \cite{Yehia 2006 a}.

The problem of constructing all integrable mechanical systems admitting a
quartic complementary integral on a Riemannian manifold has been reduced in 
\cite{yjpamast} to a single nonlinear PDE, which we call \emph{the resolving
equation}. A solution of this equation is what we call \emph{generating
function}. This generates a conservative lagrangian system together with its
quartic integral, valid on its zero-energy level. In the present article we
introduce a new solution of the resolving equation. We also formulate a
theorem that has been implicitly used in our previous work to construct
systems integrable on arbitrary energy levels from condtional ones usually
built on the zero-energy level. In its final unconditional form the
integrable system based on the new solution involves 15 parameters. Special
cases of it introduce new integrable cases of the types investigated earlier
by Toda, Bozis and others for a particle moving in the Euclidean plane. A
special case of the new system adds a rare new integrable problem in rigid
body dynamics.

\subsection{\protect\bigskip \label{1.2}Construction of integrable systems}

According to a result of Birkhoff \cite{birk}, the general natural
mechanical system (on an arbitrary 2D Riemannian configuration space) can
always be reduced in certain (isometric) coordinates $\xi ,\eta $ and time
parametrization $\tau $ to the form of a ficticious plane system described
by the Lagrangian 
\begin{equation}
L=\frac{1}{2}[\xi ^{^{\prime }2}+\eta ^{^{\prime }2}]+U,U=U(\xi ,\eta )
\label{1}
\end{equation}%
so that the equations of motion become%
\[
x^{\prime \prime }=\frac{\partial U}{\partial x}~\ \ \ \ y^{\prime \prime }=%
\frac{\partial U}{\partial y} 
\]%
are restricted to their zero-energy level 
\begin{equation}
\xi ^{\prime 2}+\eta ^{\prime 2}-2U=0  \label{2}
\end{equation}%
The energy constant $h$ for the original system enters linearly as a
parameter in the ficticious force function $U,$ which has the structure 
\begin{equation}
U=\Lambda (h-V)  \label{U}
\end{equation}%
where $V$ is the potential of the original system and $\Lambda $ is a
function that depends on the metric of the configation space.

\bigskip According to a result of \cite{Yehia 86} (see also \cite{yjpamast}%
), if an integral of motion of the mechanical system exists in the form of a
polynomial of the fourth degree in velocities, this integral can be reduced
while preserving the form of the Lagrangian (\ref{1}) to the form

\begin{equation}
I=\xi ^{\prime 4}+P\xi ^{\prime 2}+Q\xi ^{\prime }\eta ^{\prime
}+R=I_{0}(const)  \label{3}
\end{equation}%
where $R$ is given by the quadrature 
\begin{equation}
R=-\int Q\frac{\partial U}{\partial \xi }d\eta -\int [2P\frac{\partial U}{%
\partial \xi }+Q\frac{\partial U}{\partial \eta }+2U\frac{\partial Q}{%
\partial \eta }]_{0}d\xi  \label{R}
\end{equation}%
in which $[]_{0}$ means that the expression in the bracket is computed for $%
\eta $ taking an arbitrary constant value $\eta _{0}$ (say), and the other
three functions involved are expressed as 
\begin{equation}
P=\frac{\partial ^{2}F}{\partial \xi ^{2}},Q=-\frac{\partial ^{2}F}{\partial
\xi \partial \eta },U=-\frac{1}{4}(\frac{\partial ^{2}F}{\partial \xi ^{2}}+%
\frac{\partial ^{2}F}{\partial \eta ^{2}})  \label{PP}
\end{equation}%
in terms of a single generating function $F$ which satisfies the nonlinear
partial differential equation 
\begin{eqnarray}
e &=&\frac{\partial ^{2}F}{\partial \xi \partial \eta }(\frac{\partial ^{4}F%
}{\partial \xi ^{4}}-\frac{\partial ^{4}F}{\partial \eta ^{4}})+3(\frac{%
\partial ^{3}F}{\partial \xi ^{3}}\frac{\partial ^{3}F}{\partial \xi
^{2}\partial \eta }-\frac{\partial ^{3}F}{\partial \eta ^{3}}\frac{\partial
^{3}F}{\partial \eta ^{2}\partial \xi })  \nonumber \\
&&+2(\frac{\partial ^{2}F}{\partial \xi ^{2}}\frac{\partial ^{4}F}{\partial
\xi ^{3}\partial \eta }-\frac{\partial ^{2}F}{\partial \eta ^{2}}\frac{%
\partial ^{4}F}{\partial \eta ^{3}\partial \xi })  \nonumber \\
&=&0  \label{eF}
\end{eqnarray}%
The set of solutions of this equation generates all systems of the type (\ref%
{1}) having an integral of the form (\ref{3}) on the zero level of their
energy integral. Affecting all possible conformal mappings of the complex $%
\zeta =\xi +i\eta $ plane followed by a general point transformation to the
generalized coordinates $q_{1},q_{2}$ with a suitable change of the time
variable we obtain all systems of the general form\ on 2D Riemannian (or
pseudo-Riemannian) manifolds, having a quartic integral on the zero level of
their energy integral.

\subsection{ Certain forms of solutions of the resolving equation}

The fourth-order PDE (\ref{eF}) is homogeneous nonlinear of the second
degree. It is not known whether this equation is integrable, in the sense
that some procedure can be pointed out to construct all its solutions, nor
it can be reduced to an equation of lower order. The best one can do is to
try to construct solutions involving as much arbitrary parameters as
possible.

\begin{enumerate}
\item Equation (\ref{eF}) admits a solution of the form 
\begin{equation}
F=A\Phi (\xi )\Psi (\eta )  \label{ms}
\end{equation}%
where $A$ is a constant and both $\Phi $ and $\Psi $ satisfy the same
equation%
\begin{equation}
\Phi ^{(4)}(\xi )+5\frac{\Phi ^{\prime \prime }(\xi )\Phi ^{\prime \prime
\prime }(\xi )}{\Phi ^{\prime }(\xi )}-6a_{4}\Phi (\xi )=0  \label{es}
\end{equation}

in which $a_{4}$ is a separation constant. Comparing with the results of 
\cite{yjpamast} we conclude that the original isometric variables $\xi ,\eta 
$ are not practically suitable coordinates for the description of the
solution, and that a symmetric separation solution (\ref{ms}) can be more
conveniently expressed in the form%
\[
F=Apq 
\]%
in terms of the pair of variables $p,q$\ related to $\xi ,\eta $ by the
relations

\begin{eqnarray}
\xi &=&\int^{p}\frac{dp}{\sqrt[4]{a_{4}p^{4}+a_{2}p^{2}+a_{1}p+a_{0}}}, 
\nonumber \\
\eta &=&\int^{q}\frac{dq}{\sqrt[4]{a_{4}q^{4}+b_{2}q^{2}+b_{1}q+b_{0}}}
\label{co0}
\end{eqnarray}%
The three constants $a_{2},a_{1},a_{0}$ (and $b_{2},b_{1},b_{0}$) are
integration constants of the fourth-order equation (\ref{es}) for $\Phi $
(and $\Psi $). The fourth constant is eliminated by a shift of the variable $%
p$ ($q$).

\item A quite rich system, first announced in \cite{yjpamast} as the \emph{%
master }system, can be derived from the solution%
\begin{equation}
F=F_{0}(p)+G_{0}(q)+\nu pq  \label{F1}
\end{equation}%
in which%
\begin{eqnarray}
F_{0} &=&\int^{p}\frac{dp}{\sqrt[4]{%
a_{4}p^{4}+a_{3}p^{3}+a_{2}p^{2}+a_{1}p+a_{0}}}\int^{p}\frac{%
(4C_{0}+4C_{1}p+4Ap^{2}+\frac{1}{4}b_{3}p^{3})dp}{%
(a_{4}p^{4}+a_{3}p^{3}+a_{2}p^{2}+a_{1}p+a_{0})^{3/4}}  \nonumber \\
G_{0} &=&\int^{q}\frac{dq}{\sqrt[4]{%
a_{4}q^{4}+b_{3}q^{3}+b_{2}q^{2}+b_{1}q+b_{0}}}\int^{q}\frac{%
(4D_{0}+4D_{1}q+4Aq^{2}+\frac{1}{4}a_{3}q^{3})dq}{%
(a_{4}q^{4}+b_{3}q^{3}+b_{2}q^{2}+b_{1}q+b_{0})^{3/4}}  \nonumber \\
&&  \label{F0}
\end{eqnarray}%
and the relations%
\begin{eqnarray}
\xi  &=&\int^{p}\frac{dp}{\sqrt[4]{%
a_{4}p^{4}+a_{3}p^{3}+a_{2}p^{2}+a_{1}p+a_{0}}},  \nonumber \\
\eta  &=&\int^{q}\frac{dq}{\sqrt[4]{%
a_{4}q^{4}+b_{3}q^{3}+b_{2}q^{2}+b_{1}q+b_{0}}}  \label{co}
\end{eqnarray}%
generalize (\ref{co0}) by a shift of the two variables $p,q.$ This system
has the Lagrangian 
\begin{eqnarray}
L &=&\frac{1}{2}[\frac{\dot{p}^{2}}{\sqrt{%
a_{4}p^{4}+a_{3}p^{3}+a_{2}p^{2}+a_{1}p+a_{0}}}+\frac{\dot{q}^{2}}{\sqrt{%
a_{4}q^{4}+b_{3}q^{3}+b_{2}q^{2}+b_{1}q+b_{0}}}]  \nonumber \\
&&-[\frac{\nu b_{3}p^{3}+Ap^{2}+C_{1}p+C_{0}}{\sqrt{%
a_{4}p^{4}+a_{3}p^{3}+a_{2}p^{2}+a_{1}p+a_{0}}}+\frac{\nu
a_{3}q^{3}+Aq^{2}+D_{1}q+D_{0}}{\sqrt{%
a_{4}q^{4}+b_{3}q^{3}+b_{2}q^{2}+b_{1}q+b_{0}}}]  \nonumber \\
&&-\nu \lbrack \frac{q(4a_{4}p^{3}+3a_{3}p^{2}+2a_{2}p+a_{1})}{\sqrt{%
a_{4}p^{4}+a_{3}p^{3}+a_{2}p^{2}+a_{1}p+a_{0}}}+\frac{%
p(4a_{4}q^{3}+3b_{3}q^{2}+2b_{2}q+b_{1})}{\sqrt{%
a_{4}q^{4}+b_{3}q^{3}+b_{2}q^{2}+b_{1}q+b_{0}}}]  \label{ML}
\end{eqnarray}%
which involves 15 free parameters $%
a_{0},a_{1},a_{2},a_{3},a_{4},b_{0},b_{1},b_{2},b_{3},A,C_{0},C_{1},D_{0},D_{1},\nu 
$ and the quartic integral valid only on the zero level of its energy. This
integral was provided in \cite{yjpamast}. It turned out that the integral
can be reduced to somewhat simpler form%
\begin{eqnarray}
I &=&\frac{\{\dot{p}^{2}/2+\nu q(4a_{4}p^{3}+3a_{3}p^{2}+2a_{2}p+a_{1})+\nu
b_{3}p^{3}+C_{1}p+C_{0}\}^{2}}{a_{4}p^{4}+a_{3}p^{3}+a_{2}p^{2}+a_{1}p+a_{0}}%
-4\nu \dot{p}\dot{q}  \nonumber \\
&&-8\nu ^{2}\sqrt{a_{4}p^{4}+a_{3}p^{3}+a_{2}p^{2}+a_{1}p+a_{0}}\sqrt{%
a_{4}q^{4}+b_{3}q^{3}+b_{2}q^{2}+b_{1}q+b_{0}}  \nonumber \\
&&-4\nu \{\nu \lbrack
b_{2}p^{2}+a_{2}q^{2}+3pq(2a_{4}pq+b_{3}p+a_{3}q)]+2Apq+D_{1}p+C_{1}q\}
\label{MI}
\end{eqnarray}
\end{enumerate}

\section{\label{rd}Real dynamics:}

\subsection{Classes of time-parametrized geodesics.}

To construct a system that would be a priori integrable on all levels of
energy, we should receive a function $U$ from (\ref{PP}) that has the
structure (\ref{U}), i.e. involving in a linear way an arbitrary parameter $%
h $. In that case The generating function should have the form $%
F=hF_{0}+F_{1}, $ where $F_{0},F_{1}$ are functions not depending on $h.$
This is usually expected when solving linear PDEs. However, we shall rely on
solutions of the nonlinear PDE (\ref{eF}) having this property of being a
linear superposition of solutions. It is obvious from (\ref{eF}) that the
function $F$ can always be reduced to the forms $\lambda F_{0}$ ($\lambda $
arbitrary constant), but in practice we seek $F$ in the form of a linear
superposition of such expressions, so that 
\begin{equation}
F=\sum\limits_{i=1}^{k}\lambda _{i}F_{i}  \label{fup}
\end{equation}%
where $\lambda _{i},i=1,...,k$ are arbitrary parameters and $\{F_{i}\}$ do
not depend on any one of the $\lambda _{i}$s. This choice will lead to 
\begin{equation}
U=\sum\limits_{i=1}^{k}\lambda _{i}U_{i},U_{i}=-\frac{1}{4}(\frac{\partial
^{2}F_{i}}{\partial \xi ^{2}}+\frac{\partial ^{2}F_{i}}{\partial \eta ^{2}})
\label{up}
\end{equation}%
and $\{U_{i}\}$ do not depend on any one of the $\lambda _{i}$s. Each $%
\lambda _{i}$ can be regarded as the energy parameter of the real mechanical
system before reduction to the form (\ref{1}), and hence we call them
energy-like parameters. However, for a much general result we formulate the
following

\textbf{Theorem 1}

\textit{Let the system with the (reduced) Lagrangian}%
\begin{equation}
L=\frac{1}{2}[\xi ^{^{\prime }2}+\eta ^{^{\prime
}2}]+\sum\limits_{i=1}^{k}\lambda _{i}U_{i}  \label{rL0}
\end{equation}%
\textit{\bigskip in which }$()^{\prime }=\frac{d}{d\tau }$\textit{\ and }$%
\{U_{i}\}$\textit{\ do not depend on any one of the }$\lambda _{j}$\textit{,
admit on its zero-energy level an integral }%
\begin{equation}
I=I(\xi ,\eta ,\xi ^{^{\prime }},\eta ^{^{\prime }},\lambda _{1},\cdot \cdot
\cdot ,\lambda _{k})  \label{rI0}
\end{equation}%
\textit{\bigskip then the Lagrangian}%
\begin{equation}
L^{\ast }=\frac{1}{2}(\sum\limits_{i=1}^{k}\alpha _{i}U_{i})[\dot{\xi}^{2}+%
\dot{\eta}^{2}]+\frac{\sum\limits_{i=1}^{k}\beta _{i}U_{i}}{%
\sum\limits_{i=1}^{k}\alpha _{i}U_{i}}  \label{tL}
\end{equation}%
\textit{in which }$(\dot{)}=\frac{d}{dt}$\textit{\ and }$t$\textit{\ is the
natural time, admits an unconditional integral (valid for arbitrary initial
conditions) of the form}%
\begin{equation}
I^{\ast }=I(\xi ,\eta ,\Lambda \dot{\xi},\Lambda \dot{\eta},\alpha
_{1}h+\beta _{1},\cdot \cdot \cdot ,\alpha _{k}h+\beta _{k})  \label{ti}
\end{equation}%
\textit{where }$\alpha _{1},...,\alpha _{k},\beta _{1},...,\beta _{k}$%
\textit{\ are arbitrary parameters, }$\Lambda =\sum\limits_{i=1}^{k}\alpha
_{i}U_{i}$\textit{\ and }$h$\textit{\ is the energy integral for (\ref{tL}).
The energy parameter }$\ h$\textit{\ can be either substituted by its value
taken on an arbitrary motion of the system(\ref{tL}) or replaced by its
functional form in the state space}%
\[
h=\frac{1}{2}(\sum\limits_{i=1}^{k}\alpha _{i}U_{i})[\dot{\xi}^{2}+\dot{\eta}%
^{2}]-\frac{\sum\limits_{i=1}^{k}\beta _{i}U_{i}}{\sum\limits_{i=1}^{k}%
\alpha _{i}U_{i}} 
\]%
\textbf{Proof}

Introduce new arbitrary parameters $\{\alpha _{i},\beta _{i}\}$ by the
substitution $\lambda _{i}=\alpha _{i}h+\beta _{i},i=1,...,k$. The reduced
force function in (\ref{rL0}) splits into the two expressions $%
\sum\limits_{i=1}^{k}\beta _{i}U_{i}+h\sum\limits_{i=1}^{k}\alpha _{i}U_{i}.$
Now, we make the change of the independent variable $\tau $ to the original
(natural) time variable according to the relation 
\begin{equation}
d\tau =\frac{dt}{\Lambda },\Lambda =\sum\limits_{i=1}^{k}\alpha _{i}U_{i}
\label{tt0}
\end{equation}%
It is easy to see that those substitutions lead to the Lagrangian%
\[
L_{1}=L^{\ast }+h 
\]%
The additive parameter $h$ in $L_{1}$ is insignificant and discarding it
reduces $L_{1}$ to $L^{\ast }.$ The zero level of the energy integral\ for $%
L_{1}$ is the same as the $h$-level for $L^{\ast }.$

Notes:

\begin{enumerate}
\item Although we are concerned here with systems with a quartic integral,
in this theorem the integral (\ref{rI0}) is a general function of its
arguments and not necessarily quartic nor even a polynomial in velocities.

\item According to Maupertuis principle (see e.g. \cite{lan}), it should be
noted that the whole $2k-$ parameter family of mechanical systems (\ref{tL})
on their level $h$ of energy share one and the same $k-$parameter geodesic
flow (or the same $k-$ parameter family of unparametrized trajectories) on
the manifold with metric%
\begin{equation}
ds^{2}=2(\sum\limits_{i=1}^{k}\lambda _{i}U_{i})[d\xi ^{2}+d\eta ^{2}]
\label{im}
\end{equation}

\item The form (\ref{rL0}) in the independent variable $\tau $ can be
regarded as the simplest form of the $2k-$ parameter family of mechanical
systems. Let $(\xi (\tau ),\eta (\tau ))$ be the general solution of (\ref%
{rL0}). The general solution of (\ref{tL}) can be expressed in terms of the
natural time $t$ by first integrating the relation (\ref{tt0}) to get 
\begin{equation}
t=\int \Lambda d\tau =\int \sum\limits_{i=1}^{k}\alpha _{i}U_{i}(\xi (\tau
),\eta (\tau ))d\tau
\end{equation}%
and solving for $\tau $ we obtain $\tau =\tau (t),$ so that the solution of (%
\ref{tL}) is $(\xi (\tau (t)),\eta (\tau (t))).$ The form of the function $%
\tau (t)$ for each choice of the set of parameters $\{\alpha _{i}\}$
determines the way in which the mechanical system (\ref{tL}), which lives on
the -generally speaking- non-Euclidean manifold with metric%
\[
ds^{2}=2(\sum\limits_{i=1}^{k}\alpha _{i}U_{i})[d\xi ^{2}+d\eta ^{2}] 
\]%
describes its trajectory in the plane of the isometric variables $\xi ,\eta $%
.
\end{enumerate}

\section{\protect\bigskip A new solution of the resolving equation}

In the present section we try a deformation of the master system to
accomodate on more term of the product type, probably at the expense of
enforcing certain restrictions on some of the parameters figuring in (\ref%
{F0}). In fact, we assume F in the form%
\begin{equation}
F=F_{0}+K_{1}pq+K_{2}p^{2}q^{2}  \label{F}
\end{equation}

Inserting the expression (\ref{F}) in equation (\ref{eF}) and using the
relations (\ref{co}), we obtain after some manipulations a polynomial
expression in the two variables $p,q$ that must vanish identically. Those
equations have been solved under the condition that $a_{4}\neq 0,$ so that
the two polynomials that occur under the fourth degree root signs are of the
fourth degree. This resulted in only one new case, which differs from the
master system of \cite{yjpamast}. This case will be considered in detail in
the next subsections.

\subsection{\protect\bigskip The generic restricted case}

We write the Lagrangian and the comlementary integral for this case after
some transformation to more symmetric form that does not affect the
generality of the system (arbitrary shifts of the two variables will retain
the raw case):

\begin{equation}
L=\frac{1}{2}[\frac{\acute{u}^{2}}{\sqrt{au^{4}+k_{1}u^{2}+k_{0}}}+\frac{%
\acute{v}^{2}}{\sqrt{av^{4}+m_{1}v^{2}+m_{0}}}]+U  \label{L0}
\end{equation}%
\begin{eqnarray}
&&U=-\frac{N[(4av^{2}+m_{1})(4au^{4}+3k_{1}u^{2}+2k_{0})]+\nu
uv(2au^{2}+k_{1})+Ku^{2}+D}{2\sqrt{au^{4}+k_{1}u^{2}+k_{0}}}  \nonumber \\
&&-\frac{N[(4au^{2}+k_{1})(4av^{4}+3m_{1}v^{2}+2m_{0})]+\nu
uv(2av^{2}+m_{1})+Kv^{2}+E}{2\sqrt{av^{4}+m_{1}v^{2}+m_{0}}}  \label{U0}
\end{eqnarray}

\begin{eqnarray}
I &=&\frac{\{\acute{u}^{2}+N(4av^{2}+m_{1})(4au^{4}+3k_{1}u^{2}+2k_{0})+\nu
uv(2au^{2}+k_{1})+Ku^{2}+D\}^{2}}{(au^{4}+k_{1}u^{2}+k_{0})}  \nonumber \\
&&-4(8Nauv+\nu )\acute{u}\acute{v}-2(8Nauv+\nu )^{2}\sqrt{%
au^{4}+k_{1}u^{2}+k_{0}}\sqrt{av^{4}+m_{1}v^{2}+m_{0}}  \nonumber \\
&&-32aN^{2}[u^{2}(2au^{2}+k_{1})(6av^{4}+3m_{1}v^{2}+m_{0})+k_{0}v^{2}(2av^{2}+m_{1})]
\nonumber \\
&&-16Na(Eu^{2}+Dv^{2}+2Ku^{2}v^{2})  \nonumber \\
&&-4N\nu uv[8a(m_{1}u^{2}+k_{1}v^{2}+3au^{2}v^{2})+3k_{1}m_{1}]  \nonumber \\
&&-\nu ^{2}(m_{1}u^{2}+k_{1}v^{2}+6au^{2}v^{2})-4\nu Kuv  \label{I0}
\end{eqnarray}%
in which the prime represents derivative with respect to the independent
variable $\tau $ (ficticious time). Note that when the two parameters $N$
and $\nu $ vanish the Lagrangian (\ref{L0}) degenerates into a separable one
and the integral (\ref{I0}) into the square of a quadratic integral.

\bigskip The system with the Lagrangian (\ref{L0}) admits the integral (\ref%
{I0}) only on the zero-energy level of this system%
\[
\frac{1}{2}[\frac{\acute{u}^{2}}{\sqrt{au^{4}+k_{1}u^{2}+k_{0}}}+\frac{%
\acute{v}^{2}}{\sqrt{av^{4}+m_{1}v^{2}+m_{0}}}]-U=0 
\]%
This system involves 10 parameters $a,k_{0},k_{1},m_{0},m_{1},D,E,K,\nu ,N$,
of which the first five enter in both the kinetic energy and potential terms
of the Lagrangian and the last five ones enter only in the potential terms
and, moreover, they enter only linearly. The last four parameters constitute
a set of energy-like parameters (see e.g. \cite{yjpamast}). They are
essential in building the unrestricted integrable system. Comparing (\ref{L0}%
) to its counterpart in the "master" system \cite{yjpamast}, we find that (%
\ref{L0}) involves only one new parameter $N$, which is not present in the
master system. When $N$ is set equal to zero (\ref{L0}) turns out to be a
special case of the master system resulting from one restriction on the
coefficients of each of the two fourth-degree polynomials entering under the
root sign and two\ restrictions on the energy -like parameters (namely, the
vanishing of the parameters $C_{1},D_{1}$ of \cite{yjpamast}).

\subsection{\protect\bigskip Dynamics- The unrestricted generalization}

We now proceed to use those parameters to construct a general integrable
system valid on arbitrary energy level out of the restricted one.
Introducing new parameters by the relations%
\begin{eqnarray}
D &=&h_{1}-h\alpha _{1},E=h_{2}-h\alpha _{2},  \nonumber \\
K &=&h_{3}-h\alpha _{3},\nu =h_{4}-h\alpha _{4},  \nonumber
\\
N &=&h_{5}-h\alpha _{5}  \label{pars}
\end{eqnarray}%
and performing the change of independent variable to the actual-time
parametrization by using the relation

\begin{equation}
d\tau =\frac{dt}{\Lambda }  \label{tc}
\end{equation}%
where

\bigskip 
\begin{eqnarray}
\Lambda  &=&\frac{\alpha _{1}+\alpha _{3}u^{2}+\alpha
_{4}uv(2au^{2}+k_{1})+\alpha
_{5}[m_{1}u^{4}+v^{2}(4au^{4}+3k_{1}u^{2}+2k_{0})]}{\sqrt{%
au^{4}+k_{1}u^{2}+k_{0}}}  \nonumber \\
&&+\frac{\alpha _{2}+\alpha _{3}v^{2}+\alpha _{4}uv(2av^{2}+m_{1})+\alpha
_{5}[k_{1}v^{4}+u^{2}(4av^{4}+3m_{1}v^{2}+2m_{0})]}{\sqrt{%
av^{4}+m_{1}v^{2}+m_{0}}}  \label{lam}
\end{eqnarray}%
we arrive at the new Lagrangian

\begin{equation}
L=\frac{1}{2}\Lambda \lbrack \frac{\dot{u}^{2}}{\sqrt{au^{4}+k_{1}u^{2}+k_{0}%
}}+\frac{\dot{v}^{2}}{\sqrt{av^{4}+m_{1}v^{2}+m_{0}}}]-V+h  \label{LT}
\end{equation}

\begin{eqnarray}
V &=&\frac{1}{\Lambda }{\huge \{}\frac{%
h_{1}+h_{3}u^{2}+h_{4}uv(2au^{2}+k_{1})+h_{5}[m_{1}u^{4}+v^{2}(4au^{4}+3k_{1}u^{2}+2k_{0})]%
}{\sqrt{au^{4}+k_{1}u^{2}+k_{0}}}  \nonumber \\
&&+\frac{%
h_{2}+h_{3}v^{2}+h_{4}uv(2av^{2}+m_{1})+h_{5}[k_{1}v^{4}+u^{2}(4av^{4}+3m_{1}v^{2}+2m_{0})]%
}{\sqrt{av^{4}+m_{1}v^{2}+m_{0}}}{\huge \}}
\end{eqnarray}%
which admits on an arbitrary energy level $h$ the integral resulting from (%
\ref{I0}) by the substitutions (\ref{pars}) and (\ref{tc}), i.e. $\acute{u}%
\rightarrow \Lambda \dot{u},\acute{v}\rightarrow \Lambda \dot{v}$. The
integral will depend on the parameters occuring in the Lagrangian and also
on the energy constant $h.$ The last constant may be substituted by its
expression in terms of the coordinates and velocities to get the final form
free of the energy restriction. The resulting system depends on 16
parameters $a,k_{0},k_{1},m_{0},m_{1},\alpha _{1},\alpha _{2},\alpha
_{3},\alpha _{4},\alpha _{5},h_{1},h_{2},h_{3},h_{4},h_{5}$ and $h$, of
which the first nine enter in both the kinetic energy (the metric of the
configuration space) and potential terms of the Lagrangian but the last five
ones enter only in the potential.

\subsection{Special cases}

\subsubsection{Generalization of the cases of Bozis and Wojciechowski}

Let $a=k_{0}=m_{0}=1,k_{1}=m_{1}=-2.$ Under the coordinate transformation $%
p=\sin y,q=\sin x$ the Lagrangian (\ref{LT}) takes the form 
\begin{eqnarray}
L &=&\frac{1}{2}[\alpha +\beta \sin x\sin y+\gamma (2\cos ^{2}x\cos
^{2}y-\cos ^{2}x-\cos ^{2}y)+\frac{\delta _{1}}{\cos ^{2}x}+\frac{\delta _{2}%
}{\cos ^{2}y}](\dot{x}^{2}+\dot{y}^{2})  \nonumber \\
&&-\frac{a+b\sin x\sin y+c(2\cos ^{2}x\cos ^{2}y-\cos ^{2}x-\cos ^{2}y)+%
\frac{d_{1}}{\cos ^{2}x}+\frac{d_{2}}{\cos ^{2}y}}{\alpha +\beta \sin x\sin
y+\gamma (2\cos ^{2}x\cos ^{2}y-\cos ^{2}x-\cos ^{2}y)+\frac{\delta _{1}}{%
\cos ^{2}x}+\frac{\delta _{2}}{\cos ^{2}y}}  \label{gboz}
\end{eqnarray}%
When $\beta =\gamma =\delta _{1}=\delta _{2}=0$ we have, after ignoring an
insignificant additive constant 
\begin{equation}
L=\frac{1}{2}(\dot{x}^{2}+\dot{y}^{2})-[b\sin x\sin y+c(2\cos ^{2}x\cos
^{2}y-\cos ^{2}x-\cos ^{2}y)+\frac{d_{1}}{\cos ^{2}x}+\frac{d_{2}}{\cos ^{2}y%
}]  \label{boz}
\end{equation}%
This system is new. It describes plane motion of a particle in a 4-parameter
potential. The complementary integral of this system can be written as%
\begin{eqnarray}
I &=&(\dot{x}^{2}+\frac{2d_{1}}{\cos ^{2}x})(\dot{y}^{2}+\frac{2d_{2}}{\cos
^{2}y})-2\cos x\cos y(b+2c\sin x\sin y)\dot{x}\dot{y}  \nonumber \\
&&+\cos ^{2}x\cos ^{2}y(b+2c\sin x\sin y)^{2}+4c(d_{1}\cos ^{2}y+d_{2}\cos
^{2}x)
\end{eqnarray}

When $c=0$ this case reduces to a special version of that found by Bozis 
\cite{boz} and when $c=b=0$ the system becomes separable and the integral
degenerates into the product of two quadratic integrals. A slight variation
of the parameters in (\ref{gboz}) to be $k_{1}=m_{1}=2$ changes
trigonometric functions to hyperbolic (or exponential) functions, and thus
giving a new system like the type of \cite{woj1}. The analog of (\ref{boz})
gives a particle in the potential%
\begin{equation}
V=b\sinh x\sinh y+c(2\cosh ^{2}x\cosh ^{2}y-\cosh ^{2}x-\cosh ^{2}y)+\frac{%
d_{1}}{\cosh ^{2}x}+\frac{d_{2}}{\cosh ^{2}y}
\end{equation}%
with the corresponding integral%
\begin{eqnarray}
I &=&(\dot{x}^{2}+\frac{2d_{1}}{\cosh ^{2}x})(\dot{y}^{2}+\frac{2d_{2}}{%
\cosh ^{2}y})+2\cosh x\cosh y(-b+2c\sinh x\sinh y)\dot{x}\dot{y}  \nonumber
\\
&&+\cosh ^{2}x\cosh ^{2}y(b-2c\sinh x\sinh y)^{2}+4c(d_{1}\cosh
^{2}y+d_{2}\cosh ^{2}x)
\end{eqnarray}%
In a similar way, one can obtain a mix of the two types by taking $%
k_{1}=-m_{1}=2.$

\subsubsection{\protect\bigskip Systems of the Toda type}

If in (\ref{LT}) we set $a=1,k_{0}=k_{1}=m_{0}=m_{1}=0,$ the Lagrangian
takes the form

\begin{eqnarray}
L &=&\frac{1}{2}(\alpha _{0}+\alpha e^{-2x}+\beta e^{-2y}+\gamma
e^{x+y}+\delta e^{2(x+y)})(\dot{x}^{2}+\dot{y}^{2})  \nonumber \\
&&-\frac{h_{0}+ae^{-2x}+be^{-2y}+ce^{x+y}+de^{2(x+y)}}{\alpha _{0}+\alpha
e^{-2x}+\beta e^{-2y}+\gamma e^{x+y}+\delta e^{2(x+y)}}  \label{ttg}
\end{eqnarray}%
and the integral may be written, after using the energy integral to
eliminate $h,$ as%
\begin{eqnarray}
I &=&\lambda ^{4}\dot{x}^{2}\dot{y}^{2}+2\lambda ^{2}[be^{-2y}\dot{x}%
^{2}+ae^{-2x}\dot{y}^{2}+(ce^{x+y}+de^{2x+2y})\dot{x}\dot{y}]  \nonumber \\
&&+e^{2x+2y}(c+de^{x+y})^{2}+2d(be^{2x}+ae^{2y})+4abe^{-2x-2y}  \label{ttgin}
\end{eqnarray}%
where $\lambda =\alpha _{0}+\alpha e^{-2x}+\beta e^{-2y}+\gamma
e^{x+y}+\delta e^{2(x+y)}.$

A special case of the Toda type is 
\begin{equation}
L=\frac{1}{2}(\dot{x}^{2}+\dot{y}%
^{2})-(ae^{-2x}+be^{-2y}+ce^{x+y}+de^{2(x+y)})  \label{tt}
\end{equation}%
In two more special cases the configuration space degenerates into a plane.
Their lagrangians can be written in polar coordinates as done in \cite%
{yjpamast}:%
\begin{eqnarray}
L_{1} &=&\frac{1}{2}(\dot{r}^{2}+r^{2}\dot{\theta}^{2})-(Ar^{2}+\frac{B}{%
r^{2}}+\frac{Ce^{2\theta }+De^{-2\theta }}{r^{4}}) \\
L_{2} &=&\frac{1}{2}(\dot{r}^{2}+r^{2}\dot{\theta}^{2})-(\frac{%
A+Be^{-2\theta }}{r^{2}}+\frac{Ce^{\theta }}{r^{3}}+\frac{De^{2\theta }}{%
r^{4}})
\end{eqnarray}%
The last three cases of motion in the plane seem to be new, but their
potentials are not periodic in $\theta $.

\subsection{\protect\bigskip Application to rigid body dynamics}

\bigskip We now consider the general problem of motion of a rigid body about
a fixed point under the action of a combination of conservative axisymmetric
potential forces. The equations of motion for this problem can be written in
the Euler-Poisson form:

\begin{eqnarray*}
A\dot{p}+(C-B)qr &=&\gamma _{2}\frac{\partial V}{\partial \gamma _{3}}%
-\gamma _{3}\frac{\partial V}{\partial \gamma _{2}}, \\
B\dot{q}+(A-C)pr &=&\gamma _{3}\frac{\partial V}{\partial \gamma _{1}}%
-\gamma _{1}\frac{\partial V}{\partial \gamma _{3}}, \\
C\dot{r}+(B-A)pq &=&\gamma _{1}\frac{\partial V}{\partial \gamma _{2}}%
-\gamma _{2}\frac{\partial V}{\partial \gamma _{1}},
\end{eqnarray*}%
\begin{equation}
\dot{\gamma}_{1}+q\gamma _{3}-r\gamma _{2}=0,\dot{\gamma}_{2}+r\gamma
_{1}-p\gamma _{3}=0,\dot{\gamma}_{3}+p\gamma _{2}-q\gamma _{1}=0,  \label{EP}
\end{equation}%
where $A,B,C$ are the principal moments of inertia, $p,q,r$ are the
components of the angular velocity of the body and $\gamma _{1},\gamma
_{2},\gamma _{3}$ are the components of the unit vector $\mathbf{\gamma }$
fixed in space in the direction of the axis of symmetry of the force fields
applied to the body, all being referred to the principal axes of inertia at
the fixed point.

The system (\ref{EP}) admits three integrals:%
\begin{eqnarray}
I_{1} &=&\frac{1}{2}(Ap^{2}+Bq^{2}+Cr^{2})+V \\
I_{2} &=&Ap\gamma _{1}+Bq\gamma _{2}+Cr\gamma _{3}  \label{li} \\
I_{3} &=&\gamma _{1}^{2}+\gamma _{2}^{2}+\gamma _{3}^{2}=1
\end{eqnarray}

Equations (\ref{EP}) admit an equivalent representation in the Lagrangian
form (see e.g. \cite{yjpamast}), which we write here for a dynamically
symmetric body, for which $B=A$. As generalized coordinates we use the
Eulerian angles: $\psi $ the angle of precession around the axis of symmetry
of the field, $\theta $ the angle of nutation and $\varphi $ the angle of
proper rotation (about the axis of symmetry of the body). The components of
the vector $\mathbf{\gamma }$ can be expressed as 
\[
\gamma _{1}=\sin \theta \sin \varphi ,\gamma _{2}=\sin \theta \cos \varphi
,\gamma _{3}=\cos \theta 
\]
After ignoring the cyclic variable $\psi $ in the sense of Routh on the zero
level of the cyclic integral $I_{2}=0$, the Routhian of this mechanical
system expressed in the other two angles $\theta $ and $\varphi $ has the
form

\begin{equation}
R=\frac{1}{2}A[\dot{\theta}^{2}+\frac{C\sin ^{2}\theta \dot{\varphi}^{2}}{%
A-(A-C)\cos ^{2}\theta }]-V  \label{r}
\end{equation}%
Comparing the structure of this Routhian function to that of the Lagrangian (%
\ref{LT}) and recalling the procedure followed in a similar situation in 
\cite{yjpamast}, we get convinced that they become identical only in the
case of Kovalevskaya configuration $A=B=2C$.

In fact, setting $a=1,k_{1}=1,k_{0}=0,m_{1}=2,m_{0}=1,\alpha _{1}=\alpha
_{4}=\alpha _{5}=0,\alpha _{2}=\alpha _{3},$ and affecting the substitution $%
u=\frac{\cos ^{2}\theta }{2\sin \theta },v=\cos \varphi $ and renaming the
remaining parameters, we get the Lagrangian

\begin{eqnarray}
L &=&\frac{1}{2}[\dot{\theta}^{2}+\frac{\sin ^{2}\theta }{1+\sin ^{2}\theta }%
\dot{\varphi}^{2}]-V,  \label{rL} \\
V &=&2C[a\sin \theta \sin \varphi +b\sin ^{2}\theta \cos (2\varphi )+\frac{%
\lambda }{\cos ^{2}\theta }+\mu \frac{1+\sin ^{2}\theta }{\sin ^{2}\theta
\cos ^{2}\varphi }]  \label{rV}
\end{eqnarray}%
and the integral%
\begin{eqnarray}
I &=&\frac{\sin ^{6}\theta \dot{\varphi}^{4}}{(1+\sin ^{2}\theta )^{4}} 
\nonumber \\
&&+\frac{\sin ^{2}\theta \dot{\varphi}^{2}}{(1+\sin ^{2}\theta )^{2}}[2a\sin
\theta \sin \varphi +b(2-3\cos ^{2}\theta )+\frac{2\lambda \sin ^{2}\theta }{%
\cos ^{2}\theta }+\frac{4\mu }{\cos ^{2}\varphi }+\sin ^{2}\theta \dot{\theta%
}^{2}]  \nonumber \\
&&+\frac{2\dot{\theta}\dot{\varphi}\sin \theta \cos \theta \cos \varphi }{%
(1+\sin ^{2}\theta )}(a\sin \theta +2b\cos ^{2}\theta \sin \varphi )+\dot{%
\theta}^{2}[b\cos 2\varphi +\frac{2\mu }{\cos ^{2}\varphi }]  \nonumber \\
&&+\frac{4\mu ^{2}}{\sin ^{2}\theta \cos ^{4}\varphi }+\mu \lbrack \frac{%
4\lambda }{\cos ^{2}\theta \cos ^{2}\varphi }+\frac{4a\sin \varphi }{\sin
\theta \cos ^{2}\varphi }+2b(2\sin ^{2}\theta +\frac{1-3\sin ^{2}\theta }{%
\sin ^{2}\theta \cos ^{2}\varphi })]  \nonumber \\
&&+(\frac{1}{2}a^{2}-2ab\sin \theta \sin \varphi )(1-2\sin ^{2}\theta \cos
^{2}\varphi )+\frac{1}{2}b^{2}\sin ^{2}\theta \lbrack \sin ^{2}\theta (\cos
4\varphi -1)+4]  \nonumber \\
&&+2\lambda b\tan ^{2}\theta \cos 2\varphi   \label{rI}
\end{eqnarray}%
\qquad \qquad \qquad \qquad \qquad \qquad 

To facilitate comparison with other results, we now express the last
integrable case of rigid body dynamics in the Euler-Poisson variables, i.e.
as a solution of the system (\ref{EP}), in the next\bigskip

\bigskip \textbf{Theorem 2:}

\textit{For a rigid body with moments of inertia satisfying }$A=B=2C$\textit{%
\ and for the potential}%
\begin{equation}
V=2C[a\gamma _{1}+b(\gamma _{1}^{2}-\gamma _{2}^{2})+\frac{\lambda }{\gamma
_{3}^{2}}+\delta \frac{2-\gamma _{3}^{2}}{\gamma _{2}^{2}}]  \label{V}
\end{equation}%
\textit{equations (\ref{EP}) are integrable on the level }$I_{2}=0$\textit{.
The complementary integral has the form}%
\begin{eqnarray}
I &=&(p^{2}-q^{2}-a\gamma _{1}+b\gamma _{3}^{2}-\frac{\lambda (\gamma
_{1}^{2}-\gamma _{2}^{2})}{\gamma _{3}^{2}})^{2}+(2pq-a\gamma _{2}-\frac{%
2\lambda \gamma _{1}\gamma _{2}}{\gamma _{3}^{2}})^{2}  \nonumber \\
&&+\frac{\delta }{\gamma _{2}^{2}}[2(p^{2}+q^{2})\gamma _{3}^{2}-2a\gamma
_{1}\gamma _{3}^{2}-2\lambda \gamma _{1}^{2}+2b+\frac{\delta \gamma _{3}^{4}%
}{\gamma _{2}^{2}}]  \label{i}
\end{eqnarray}

This case is new. For comparison we provide a table of presently known
integrable potentials related to the type (\ref{V}), which admit a quartic
integral under the condition $A=B=2C$:

\begin{center}
\begin{tabular}{ll}
Author- year & Potential \\ 
Kovalevskaya \cite{kov}1889 & $V_{1}=a_{1}\gamma _{1}+a_{2}\gamma _{2}$ \\ 
Chaplygin \cite{chap} 1903 & $V_{2}=b_{1}(\gamma _{1}^{2}-\gamma
_{2}^{2})+b_{2}\gamma _{1}\gamma _{2}$ \\ 
Goriatchev\cite{gor16} 1916 & $V_{3}=a\gamma _{1}+a_{2}\gamma _{2}+b(\gamma
_{1}^{2}-\gamma _{2}^{2})+b_{1}\gamma _{1}\gamma _{2}+\frac{\lambda }{\gamma
_{3}^{2}}$ \\ 
Yehia \cite{yjpamast} 2006 & 
\begin{tabular}{l}
$V_{4}=b(\gamma _{1}^{2}-\gamma _{2}^{2})+\frac{\lambda }{\gamma _{3}^{2}}%
+\rho (\frac{1}{\gamma _{3}^{4}}-\frac{1}{\gamma _{3}^{6}})+(2-\gamma
_{3}^{2})(\frac{\nu }{\gamma _{1}^{2}}+\frac{\delta }{\gamma _{2}^{2}})$ \\ 
$V_{5}=a\gamma _{1}+\frac{\lambda }{\gamma _{3}^{2}}+\frac{\varepsilon }{%
\sqrt{\gamma _{1}^{2}+\gamma _{2}^{2}}}+\frac{(2-\gamma _{3}^{2})}{\gamma
_{2}^{2}}[\delta +\mu \frac{\gamma _{1}}{\sqrt{\gamma _{1}^{2}+\gamma
_{2}^{2}}}]$%
\end{tabular}%
\end{tabular}
\end{center}

\bigskip The potential (\ref{V}) involves a collection of parameters: $a$ of
the Kovalevskaya type, $b$ of the Chaplygin type, the Goriatchev parameter $%
\lambda $ and the parameter $\delta $ figuring in both new cases announced
in our work \cite{yjpamast}, but the combination (\ref{V}) is new.

\end{document}